\documentclass[]{spie}  

\usepackage{amsmath,amsfonts,amssymb}
\usepackage{graphicx}
\usepackage[colorlinks=true, allcolors=blue]{hyperref}

\title{ANTARES: Progress towards building a `Broker' of time-domain alerts}

\author[a]{Abhijit~Saha}
\author[b]{Zhe~Wang}
\author[a]{Thomas~Matheson}
\author[b]{Gautham~Narayan}
\author[b]{Richard~Snodgrass}
\author[b]{John~Kececioglu}
\author[b]{Carlos~Scheidegger}
\author[c]{Tim Axelrod}
\author[d]{Tim Jenness}
\author[a]{Stephen Ridgway}
\author[e]{Robert Seaman}
\author[b]{Clark Taylor}
\author[b]{Jackson Toeniskoetter}
\author[b]{Eric Welch}
\author[b]{Shuo Yang}
\author[f]{Tayeb Zaidi}

\affil[a]{NOAO, 950 N Cherry Avenue, Tucson, AZ 85719 USA}
\affil[b]{Department of Computer Science, University of Arizona, Tucson, AZ 85721 USA}
\affil[c]{Steward Observatory, University of Arizona, Tucson AZ 85719 USA}
\affil[d]{LSST Project Office, 950 N Cherry Avenue, Tucson AZ 85719 USA}
\affil[e]{Lunar \& Planetary Laboratory, University of Arizona, Tucson, AZ 85719 USA}
\affil[f]{Department of Physics \& Astronomy, Macalester College, Saint Paul, MN 55105 USA}

\authorinfo{Further author information: (Send correspondence to A. Saha)\\ A.Saha: E-mail: saha@noao.edu, Telephone: +1 520 318 8288}

\begin{document} 
\maketitle

\begin{abstract}
The Arizona-NOAO Temporal Analysis and Response to Events System
(ANTARES) is a joint effort of NOAO and the Department of Computer
Science at the University of Arizona to build prototype software to
process alerts from time-domain surveys, especially LSST, to identify
those alerts that must be followed up immediately.  Value is added by
annotating incoming alerts with existing information from previous
surveys and compilations across the electromagnetic spectrum and from
the history of past alerts. Comparison against a knowledge repository
of properties and features of known or predicted kinds of variable
phenomena is used for categorization. The architecture and algorithms
being employed are described.

\end{abstract}


\section{INTRODUCTION}
\label{sec:intro}  
The Arizona-NOAO Temporal Analysis and Response to Events System
(ANTARES) is a joint project of the National Optical Astronomy
Observatory and the Department of Computer Science at the University
of Arizona. The goal is to build the software infrastructure necessary
to process and filter alerts produced by time-domain surveys, with the
ultimate source of such alerts being the Large Synoptic Survey
Telescope (LSST)\cite{2008arXiv0805.2366I,2014htu..conf...19K,2016SPIE-Kahn}.  
The conceptual design was laid out in a previous
SPIE proceeding~\cite{saha14}.  The primary goal of ANTARES is to
recognize `interesting' alerts that are uncharacteristic of known
kinds of variables, so that follow-up observations for further
elucidation (which are likely to be time-critical) can be done as soon
as permissible.  To do this, the ANTARES broker adds value to alerts
by annotating them with information from external sources such as
previous surveys from across the electromagnetic spectrum and then
comparing against current knowledge of other astronomical
sources. These annotated alerts are stored, so the temporal history of
any past alerts at a given sky location provides further elucidation
for analysis.  The key discriminator is a {\it Touchstone}, which is a
knowledge repository of properties and features of known or predicted
kinds of variable astronomical sources. An incoming alert's features
can be compared to the information in the Touchstone through a series
of filtering stages to ascertain whether it is `interesting'. For the
prototype, `interesting' is defined as the rarest or most unusual
alert or alerts that need immediate follow-up observations; the
architecture supports incorporating alternate logic and goals. The
system is designed to be flexible: customized filters may be inserted  
at multiple points throughout the process flow to enable specific time-critical use cases 
(which may change over the duration of the survey, requiring re-configuration of such filtering stages).
The repository of annotated alerts, including
association with known objects in the sky, is an additional data
product that can serve the needs of users seeking categories of time
varying phenomena that are not critical for immediate follow up.

An
architectural design consistent with these goals and scope was
presented in reference~\citenum{saha14}. Since then, a functioning
software framework has been erected and initial filtering stages have
been constructed. Tools and algorithms that will be needed for the
construction of later filtering stages have been, and continue to be,
developed.  Preliminary implementation of external astronomical object
catalogs is in place and the Touchstone is being populated.  In this
paper we describe the current state of development of these elements
of ANTARES, with discussion of the architecture (and its operational
stability and scalability), along with examples of some working
filtering algorithms.
 
\section{System Architecture}
\label{sec:arch}

ANTARES is a distributed computing system running on a
cluster of multiple machines. The system experiences very different
workloads during day (when LSST is not in operation) and night (when
LSST is in operation).

\begin{figure} [!ht]
 \includegraphics[height=20cm]{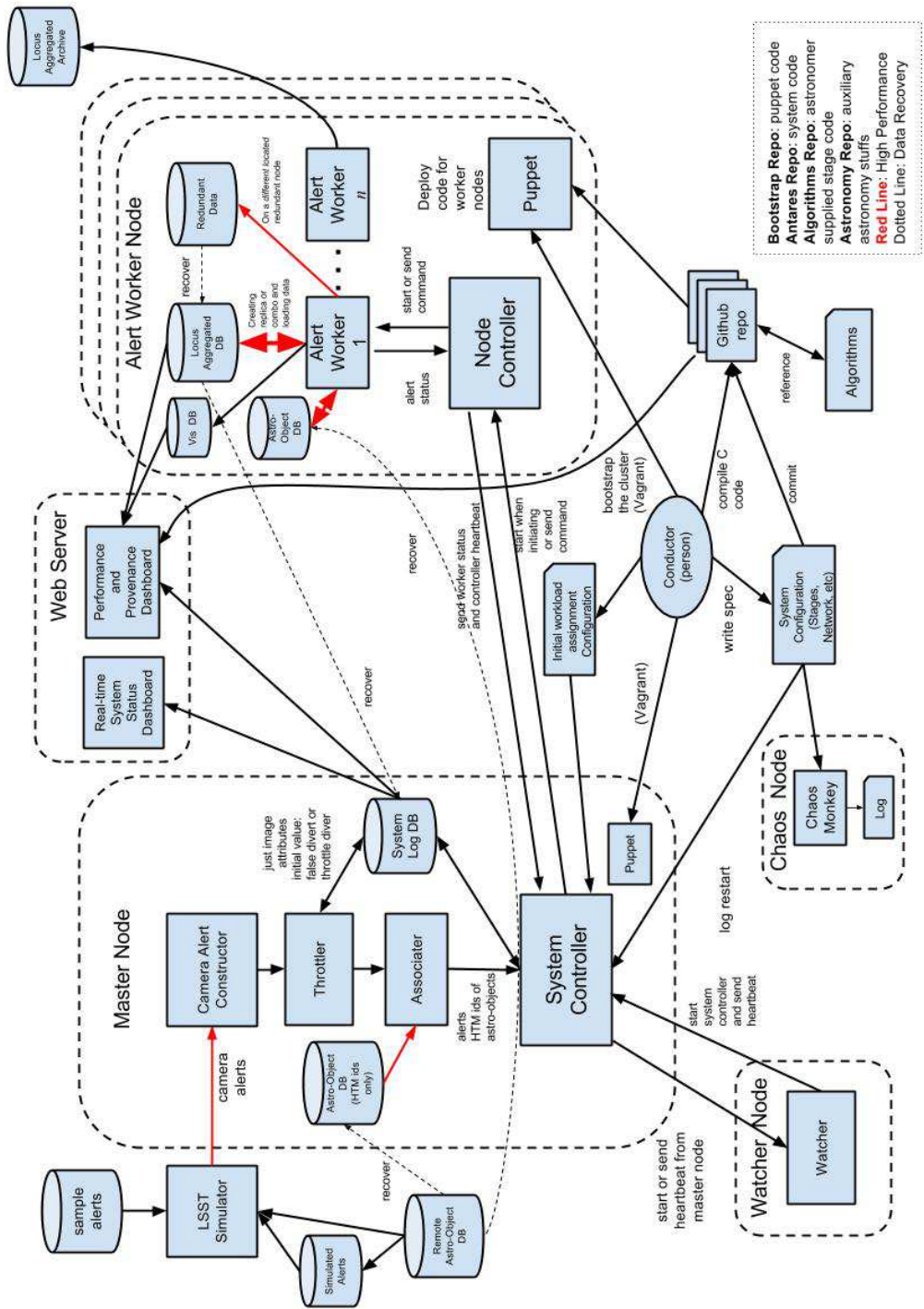}
 \caption[] {Schematic diagram for ANTARES system architecture \label{fig:arch1} }
 \end{figure}

Figure~\ref{fig:arch1} shows the ANTARES logical system architecture.
In this architecture, physical machines are denoted by
dashed lines: the {\it Master Node}, multiple {\it Alert Worker
  Nodes}, a {\it Watcher Node}, and a {\it Chaos Node} (more on each
below), all physically located in the cluster, as well as a (possibly
remote) {\it Web Server}.

During the day, a person called the {\it Conductor} prepares the {\it
  workload assignment configuration}, including deciding how many
stages to include, adjusting the time allotment for each stage, and specifying
which stage code to use that night (a separate committee vets stage
code submitted by astronomers). The Conductor also launches the system
manually in the late afternoon, with a {\it Puppet instance}\cite{Puppet2015} running
on each node that extracts the appropriate code from a {\it GitHub}
repository,
based on the configuration for that night.

At night, ANTARES receives alerts from LSST and processes them on the
Master/Alert Worker Node cluster. Then, after throttling (diverting alerts that
overflow designed ANTARES capacity), the alerts are each associated with the
AstroObject(s) located close to them in the sky (this association uses SciSQL to make a single SQL query to the database, handling all the alerts at once). The
alerts are then sent to different worker nodes. Each worker node could
have multiple {\it Alert Workers}, on which the alerts will be
processed. On each Alert Worker Node is also a portion of the {\it
  Astro-Object Database}, a portion of the {\it Locus-Aggregated
  Database}, and a {\it Visualization Database}. All of the Alert Workers
run the same alert processing pipeline, which contains multiple 
stages, each making different decisions or calculations. Some examples of 
algorithms executed in the various stages are discussed below in \S~\ref{sec:filtering}. 
Fortunately, the stages for
individual camera (or incoming) alerts can be processed entirely in
parallel, within the Alert Worker Node. The final results (the
rarest-of-the-rare camera alerts, each with derived properties stating
why the system characterized that alert as such) will be collected by
the Master Node and published by sending them to various external Alert
Brokers, not shown.

ANTARES ensures that alerts are stored on at least two Alert Worker
Nodes to ensure one-node failure resiliency (as will be discussed
shortly). Each camera alert, alert replica (created to examine various
scenarios in parallel), and alert combo (collections of alert
replicas, created to analyze various {\it combined} scenarios) is
stored in multiple places ({\it Redundant Data}).  Finally,
distributed logs are maintained to keep track of the progress of the
processing, at node and camera alert granularities.

During the following daytime, the system will perform data processing
and data migration (from the individual Alert Worker Node
Locus-Aggregated Database instances) to aggregate all the generated
data from the previous night and store it in the {\it Locus-Aggregated
  Archive}.
 
ANTARES has implemented single-node failure resilience, which means if
any one node fails, whether it is the Master Node, a Worker Node, or
one of the other nodes shown, the system’s functionality won't be
affected (though it may be that computations on the alerts then in
transit may be lost). Furthermore, the failed node will be restarted
after the problem has been detected by the system. In order to
implement single node failure resilience, an additional node called
{\it Watcher} is included. Each Alert Worker node will send heartbeats
to the Master Node.  At runtime, if an Alert Worker Node fails, the
Master Node will detect this and restart that node. The Master Node
and the Watcher Node will send heartbeats to each other. If one of
them fails, the counterpart will detect this and restart the failed
node. We help ensure robustness of failure resilience by introducing a
{\it Chaos Monkey Node}~\cite{bennett12,Tseitlin:2013:AO:2492007.2492022} into the system. This node
randomly kills one of the nodes (including perhaps itself), presumably
triggering restart actions, and repeating within a certain time
interval. Our plan is to run Chaos Monkey in all tests to uncover
difficult bugs. The expectation is that extant failure resilience bugs
would be triggered during initial testing of the system so that they
can be fixed earlier. This helps ensure the system achieves adequate
robustness.

ANTARES also provides visualization tools to the user. There are two
visualizations, for different purposes. One is the {\it Real-time
  System Status Dashboard}, which the current status of
the system as it runs during the night. From this dashboard, the user can see how many
alerts are being processed, the status of each node and process,
etc. The other visualization is the {\it Performance and Provenance
  Dashboard}. This dashboard visualizes after the fact, i.e., in the
days and weeks after a nightly run, aggregated logging information
generated during that night. The purpose of this dashboard is to help
users and the Conductor analyze the performance of ANTARES and also
easily locate and understand provenance of derived properties (by
showing the actual code used to compute a derived property, the inputs
to that code, the code that computed those inputs, and on up the
processing stream).


\section{Object Association}
\label{sec:assoc}

An incoming alert comes with a position on the sky that is matched
against existing catalogs of astrophysical objects (astro-objects) in
the Astro-Objects Database.  It is also matched against any past
alerts at this location that may be in the Locus Aggregated DB, which
contains information about any alerted behavior in the past, along
with decisions that ANTARES may have made, given the information
available with past alerts.  These associations furnish value-added
features (e.g., time-scale/period, amplitude, color, whether X-ray
source or radio source, proximity on the sky to a known galaxy, etc.)
to the current alert, that can then be used in the filtering stages
described below.  
  
Object association with the AstroObject Database is performed by calculating 
each object's 20 digit index on a hierarchical triangular mesh (HTM)\cite{2001misk.conf..631K} 
used to cover the sphere\cite{Szalay2007}. 
The quad-tree search of the HTM is implemented in SciSQL. 
Association radii are different for the various 
astronomical catalogs, as the resolution of surveys in different regions of the 
electromagnetic spectrum can vary by orders of magnitude.
The compilation of a comprehensive Astro-Objects Database is currently in progress.

\section{Filtering Algorithms}
\label{sec:filtering}

The initial ANTARES implementation allows us to begin testing various
filtering algorithms to find which ones are most efficacious.  Our
primary goal is to find alerts that are somehow different from those
that are triggered by objects we know.  The Touchstone is the
repository of features from known kinds of variable phenomena or of
models of phenomena that are expected to exist, but have not been seen
before.  The astrophysical objects (astro-objects) that populate the
Touchstone have labels that show the category/class of astro-objects
they represent.  Each filtering stage, depending on its function,
utilizes some subset of all the available features to ascertain
whether the alert under consideration, or the history of alerts of
that source taken together, could come from one or other class of
phenomenon represented in the Touchstone.  Our primary goal is to
identify alerts that are \emph{not likely} to have come from known
classes of astro-objects, or those that are likely to have come
\emph{only} from predicted but as yet unseen astro-objects, based on
their \emph{modeled features} in the Touchstone.  In the following
sub-sections, we discuss some of the filtering stages we have
implemented.

\subsection{Variability Probability Distribution Function}
\label{sec:vpdf}

Many alerts from time-domain surveys such as LSST will contain little
information.  If this is the first alert at a specific location on the
sky, then the alert will essentially be only a reference magnitude, a
change in magnitude, and the position on the sky.  Ancillary data from
other catalogs could help a great deal, as described elsewhere in this
document, but, in the absence of any external information at that locus, there is
still a process to assess how unusual an alert might be.  Knowing the
position of the alert on the sky tells us where it is in Galactic
coordinates.  Given an estimate of the range of variability for
stellar sources, one can use the anticipated stellar population along
a given Galactic line of sight to decide what amount of change in
brightness is interesting.

There are several time-domain studies of Galactic variables, but few
that have consistent measurements of all types of stars, not just
known variables.  The stars observed by the \emph{Kepler}
spacecraft~\cite{Borucki10} represent such a sample.  \emph{Kepler}
uses a 30-minute cadence over 3-month-long quarters on a wide variety
of stars.  We use the Q6 data set selected and filtered as discussed
by reference~\citenum{Ridgway14}.  The final sample contains 155,347
stars.  There are many broad classes of stars missing from the sample,
such as white dwarfs.  In addition, the length of coverage misses
variability on longer time scales.  Nonetheless, there is no better
sample of variability over a broad range of stellar types currently
available. The \emph{Gaia} mission~\cite{Perryman01} will provide a larger
and more complete sample.

For each star in the \emph{Kepler} sample, we define a clipped mean
and calculate the distribution of deviations from that mean.  (We use
a clipped mean so that eclipsing objects have a mean value that
represents the non-eclipsed brightness.)  We characterize the
deviation as a magnitude change so that it is a relative measurement.
We then resample each epoch of deviations using the error of the
measurements in order to more robustly estimate the potential range of
deviation.  This produces a slight but noticeable increase in the
range.

\begin{figure*}[tb]
\centering
\includegraphics[width=0.95\textwidth]{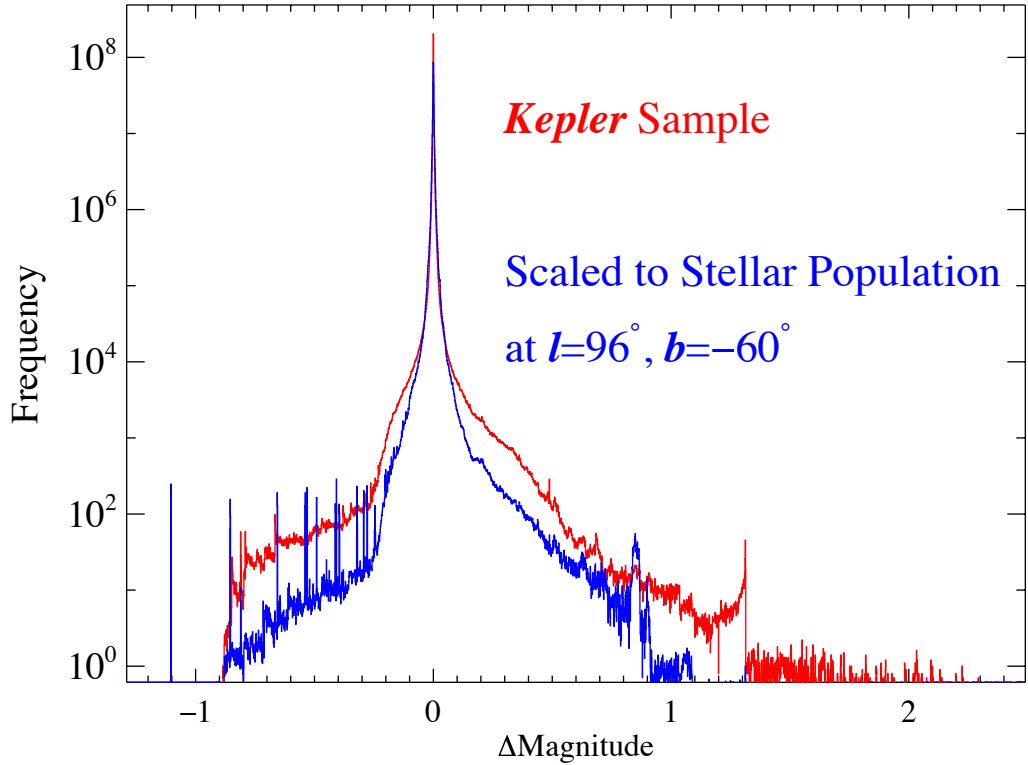}
\caption{Frequency of variability derived from the \emph{Kepler}
  observations (\emph{blue}).  The frequency distribution in
  \emph{red} shows the mapping from the stellar distribution in the
  \emph{Kepler} field to a 10 square degree field centered at
  $l=90^{\circ}$, $b=-60^{\circ}$ (the nominal LSST field).  The
  variability distribution is similar, but shows some differences that
  result from the relative stellar content of the two fields.  Note
  that frequency is on a log scale.}
\label{fig:vpdf}
\end{figure*}

The stars observed with \emph{Kepler} were studied
beforehand~\cite{Brown11} to characterize their effective temperature
($T_{eff}$) and surface gravity (log~g).  These values were updated
with more accurate estimates as reported in reference~\citenum{Huber14}.  Using these stellar parameters, we can sort the
\emph{Kepler} sample into bins using temperature as a proxy for
stellar type and log~g to separate dwarfs from giants.  We can then
create a variability distribution for each bin, where the bin sizes
were selected to ensure a reasonable number of stars in each.  We
identify this as a variability probability distribution function.  We
then used the Besan\c{c}on Galaxy model~\cite{Robin03} to predict the
distribution of stars in an LSST image for lines of sight throughout
the Galaxy.  The model includes $T_{eff}$ and log~g for each star.  We
used a variable step size for sampling the Galaxy, ranging from
5$^{\circ}$ to 20$^{\circ}$ depending on the stellar density.  We then
smoothly interpolate between these samples to get a distribution for
any particular pointing.  Areas close to the plane (within $\sim$
10$^{\circ}$) are poorly simulated as extinction is highly variable.
We take the variability distributions derived from the \emph{Kepler}
observations and map them onto the stellar population as predicted by
the Besan\c{c}on model to create a model of variability for that
particular LSST pointing.  Figure~\ref{fig:vpdf} illustrates this
transformation.

\subsection{Utilizing the History of Alerts in Locus Aggregated DB}
We expect that the majority of alerts will come from astro-objects
that are repeat variables within the Galaxy with common features. As
the LSST survey progresses, such sources will have multiple past
alerts. When a new alert from such an astro-object arrives,
association with past alerts in the Locus Aggregated DB will quickly
identify them as run-of-the-mill variables, thus allowing a large
fraction of alerts to be quickly diverted.  For those that are repeat
offenders, but have not yet been reconciled as common variables,
multiple measures of brightness at different epochs (and in different
passbands) become available as the survey progresses.  Various types
of analyses then become possible.

\subsection{The Touchstone and its Use: some early examples}
\label{sec:touchstone}
As described above, the Touchstone is a repository for useful features
of known kinds of phenomena, be they regular variables or
transients. They can also contain models of predicted kinds of
astro-objects or phenomena, which have rarely if ever been actually
seen due to their rarity.  To keep ANTARES operating efficiently, the
set of features needs to be manageable. This calls for wide
experimentation to see what set of features offer maximum leverage in
sorting between alerts from different kinds of source phenomena.
In practice, it is more efficient to have several Touchstones, each holding those features
that are used for the particular filtering stage(s) it serves, even though here we refer to it in the singular. 
Populating the Touchstone from a comprehensive set of known variables
of all types that offer the most efficacious testing of alerts is a
key scientific pursuit in the construction of ANTARES.  We describe
here an early example.

The most common feature currency that is self-supplied by LSST and so
available for all astro-objects are the observations themselves,
i.e., observation epoch $t$ and observed magnitude $m$ in multiple
bands. We refer to the set of observed $t$ and $m$ as the `light
curve,' disambiguated from the folded light curve of a periodic
variable, which we will call the `periodic light curve.'  There is
thus clear motivation to construct and examine features out of the light curve.

In an early experiment the time domain observation of variables from
the LINEAR data set~\cite{Palaversa13} were examined.  Folded
period light curves of well-labeled periodic variables in this data
set were chosen and splines were fitted to obtain a smooth
representation thereof.  The spline parameters were treated as
features, and subjected to a Principal Component Analysis (PCA).
Visualization techniques were then applied to see if and how the
projections on different planes of the PCA eigenvalue hyperspace for
the sample objects separate them according to the classification
labels.  A demonstration of the success of this effort and of the
visualization of results is made in the oral presentation.  This
method is also extendable to light curves of transients, where there
is no folding by period required and a spline is fitted to the light
curve.

The LINEAR data set offers only one passband, but the method can
clearly be extended to multiband light curves.  Efforts are ongoing
to incorporate other data sets with multiband light curves and on
transient light curves, such as of different classes of supernovae and
novae.

To utilize this approach, however, one will have to wait during the
survey for enough observations to be made to define a usable light
curve.  A pressing issue is that of gleaning more with just a few
observations.  In fact, this is critical for being able to tell apart
unusual events when they first appear, and which are short-lived
transients. The main thrust of that effort must proceed accordingly.

An unknown source on which we need to make a quick decision offers a
minimal set of ${t,m}$ (epochs and magnitudes) in any band. If we have
$n$ such observations we get $n(n-1)/2$ distinct pairs of
measurements, for each of which we get a set of $\Delta t$ and $\Delta
m$.  Now consider a well defined class of object (or collection of
well defined classes of objects), for each of which we have a large
set of ${\Delta t, \Delta m}$.  If the sampling of all $t$'s is random,
then the two dimensional histogram of all measurements of the sample
objects in the ${\Delta t, \Delta m}$ plane is an empirical
probability surface drawn from all kinds of objects included in the
sample.  It provides a way to test a single incoming alert (which
arises from the difference of two measurements spaced in time) in that
we can read from our histogram the relative probability that the given
single measured ${\Delta t, \Delta m}$ could come from an astro-object
belonging to the set of astro-objects that have defined the
probability surface above.  If we have additional points for the light
curve of the alerted objects, say a total of $j$ measurements, then we
can multiply the $j(j-1)/2$ individual relative probability values on
the ${\Delta t, \Delta m}$ test surface and infer the likelihood that
this set of measurements could be drawn from the set of astro-objects
defining the probability surface.  This is easily extendable to
multiple passbands.  The likelihoods in individual bands need only to
be multiplied together.

We have had some initial success in implementing this approach, with
positive results. The compilation of a comprehensive probability
surface presents many (not insurmountable) challenges that are labor
intensive and is an ongoing investigation.

\section{Developments Planned for the Near Future}
We are in the process of transitioning to operation on a cluster that fully implements 
the architectural design. The goal that we expect to achieve over the next 2 years
is to build a fully functional machine that works effectively on live surveys in the pre-LSST 
era.  Ideally, to expand the proto-type to handle LSST alert volumes  we should 
only have to expand the hardware, mainly the number of alert worker nodes. 

Here are a few of the specific items we will be working on:
\begin{enumerate}

\item
Complete the implementation of the architecture, specifically the Redundant Data store to recover alerts in transit, the VisDB, and the External Locus-Aggregate Archive.

\item
Add significant functionality to the dashboard.

\item
Replace the relational Locus-Aggregated Alert DB with a bespoke, much higher performance store, to contend with extremely high data rates.

\item
Expand the Touchstone, and proceed towards populating its various instances with relevant features.  Examine the efficacy of features: experiment to find the ones that give most purchase on filtering.

\item
Implement more filtering stages and evaluate their efficacy, especially to exercise the alert replication and combo functionality with realistic use cases.  Test on pseudo alerts generated from time-domain data-sets already analyzed, since these also inform us about how well we are doing.

\item 
Continually curate the Touchstone (including its various instances) to get greater completeness that represents  all known classes of variable phenomena.

\item
Transition to working on live streams, such as from the Catalina Sky Survey and the Palomar Transit Factory.

\end{enumerate}

\section{Summary}

We have described the architecture of ANTARES, whose implementation is proceeding apace. This provides for us the scaffolding to test algorithms, weed out the ineffective ones, and sharpen the ones that show good promise. This will be the bulk of the scientific endeavor in the next few years.  We plan on holding an external review of the project in late 2016, followed by meetings and workshops with the user community to disseminate what the community can expect out of ANTARES; begin to organize follow-up of time domain triggers around ANTARES functionality; and to ingest the desires and expectations of potential users. 

\acknowledgements
We acknowledge the NSF INSPIRE grant (CISE AST-1344204, PI:Snodgrass)
that supports this work.

\bibliography{antares2} 
\bibliographystyle{spiebib} 

\end{document}